\documentclass[final,3p,times,twocolumn]{elsarticle}
\usepackage{amssymb}
\usepackage{amsmath}
\usepackage{color}
\usepackage{graphicx}
\usepackage[english]{babel}

\journal{Physics Letters B}

\begin{document}

\begin{frontmatter}



\title{Evidence for $\Delta(2200)7/2^-$ from photoproduction \\and
consequence for chiral-symmetry restoration at high mass}


\author[HISKP,GATCHINA]{A.V.~Anisovich}
\author[JLAB]{V.~Burkert}
\author[HISKP]{J.~Hartmann}
\author[HISKP]{E.~Klempt}
\author[HISKP,GATCHINA]{V.A.~Nikonov}
\author[JLAB]{E.~Pasyuk}
\author[HISKP,GATCHINA]{A.V.~Sarantsev}
\author[USC]{S.~Strauch}
\author[HISKP]{U.~Thoma}

\address[HISKP]{Helmholtz--Institut f\"ur Strahlen-- und Kernphysik, Universit\"at Bonn, 53115 Bonn, Germany}
\address[GATCHINA]{NRC ``Kurchatov Institute'', Petersburg Nuclear Physics Institute, Gatchina, 188300 Russia}
\address[JLAB]{Thomas Jefferson National Accelerator Facility, 12000 Jefferson Avenue, Newport News, Virginia, USA}
\address[USC]{Department of Physics and Astronomy, University of South Carolina, Columbia, South Carolina, USA}

\begin{abstract}
We report a partial-wave analysis of new data on the double-polarization variable $E$ for the reactions $\gamma p\to \pi^+ n$ and $\gamma p\to \pi^0 p$ and of further data published earlier. The analysis within the Bonn-Gatchina (BnGa) formalism reveals evidence for a poorly known baryon resonance, the one-star $\Delta(2200)7/2^-$. This is the lowest-mass $\Delta^*$ resonance with spin-parity $J^P=7/2^-$. Its mass is significantly higher than the mass of its parity partner $\Delta(1950)7/2^+$ which is the lowest-mass $\Delta^*$ resonance with spin-parity $J^P=7/2^+$. It has
been suggested that chiral symmetry might be restored in the high-mass region of hadron excitations, and that
these two resonances should be degenerate in mass. Our findings are in conflict with this prediction.
\end{abstract}

\begin{keyword}
Partial-wave analysis, baryon resonance, photoproduction, chiral symmetry



\end{keyword}

\end{frontmatter}



SU(3) symmetry and the conjecture that mesons and baryons are
composed of constituent quarks \cite{GellMann:1964nj,Zweig:1964jf}
paved the path to an understanding of the particle zoo.  A
constituent light-quark mass of about 350~MeV was required to
reproduce the masses of ground-state baryons; the
$N$--$\Delta(1232)$ mass splitting and the pattern of negative- and
positive-parity excited baryons was interpreted as an effect of a
QCD hyperfine interaction between these constituent quarks
\cite{Isgur:1977ef,Isgur:1978xj}. However, low-energy approximations
of QCD \cite{Weinberg:1978kz} lead to the Gell-Mann-Oakes-Renner
relation \cite{GellMann:1968rz} which assigns a mass of a few MeV to
light (current) quarks. The mass gap between current and constituent
quarks is interpreted by spontaneous breaking of the chiral symmetry
expected for nearly massless quarks
\cite{Nambu:1960tm,Goldstone:1961eq}. An important consequence is
the large mass gap between chiral partners: the masses of the $\rho(770)$ meson
with spin-parity $J^P=1^-$ and its chiral partner
$a_1(1260)$ with $J^P=1^+$ differ by about
500\,MeV, those of the $J^P=1/2^+$
nucleon and its negative-parity partner $N_{1/2^-}(1535)$ by about 600~MeV.

In spite of these large mass shifts between low-mass parity partners, mesons and baryons
at higher masses are often observed in parity doublets, in pairs of resonances having the same total spin $J$, opposite parities, and about the same mass. There are, e.g., four positive-parity and four negative-parity $\Delta^*$ resonances at about 1900~MeV:\\[-3ex]
\begin{center}
{\small
\begin{tabular}{cccc}
$\Delta(1910)1/2^+$ \hspace{-3mm}&\hspace{-3mm} $\Delta(1920)3/2^+$ \hspace{-3mm}&\hspace{-3mm} $\Delta(1905)5/2^+$ \hspace{-3mm}&\hspace{-3mm} \boldmath$\Delta(1950)7/2^+$\phantom{.}\\
$\Delta(1900)1/2^-$ \hspace{-3mm}&\hspace{-3mm} $\Delta(1940)3/2^-$ \hspace{-3mm}&\hspace{-3mm} $\Delta(1930)5/2^-$ \hspace{-3mm}&\hspace{-3mm} \boldmath$\Delta(2200)7/2^-$\,.
\end{tabular}\vspace{-1mm}}
 \end{center}

The resonances with $J^P=1/2^\pm, 3/2^\pm, 5/2^\pm$ are nearly {\it mass-degenerate}, in particular when their natural widths of a few hundred MeV are considered. This example and similar observations in mesons and baryons have led to the conjecture that chiral symmetry might be effectively restored in highly excited hadrons \cite{Glozman:1999tk,Glozman:2003bt} and has stimulated a vivid discussion; we quote here a few recent reviews \cite{Jaffe:2004ph,Glozman:2007ek,Afonin:2007mj,Shifman:2007xn}. Based on the hypothesis of chiral-symmetry restoration, the parity partner of $\Delta(1950)7/2^+$ with spin-parity $7/2^-$ should also have a mass of about 1900 or 1950\,MeV while quark models \cite{Loring:2001kx} and AdS/QCD \cite{Forkel:2008un} predict $\approx$2200\,MeV. The Review of Particle Properties lists the one-star $\Delta(2200)7/2^-$ resonance and thus disfavors models assuming that chiral symmetry could be restored in high-mass baryons. But clearly, a resonance for which the evidence for its existence is estimated to be poor (the definition of the one-star rating) cannot decide on this important issue. It is essential to refute or confirm the existence of this state.

The 250-MeV mass splitting between $\Delta(1950)7/2^+$ and $\Delta(2200)7/2^-$ -- unexpected when chiral symmetry is effectively restored in highly excited hadrons -- points to a more general
concern: resonances falling onto a leading Regge trajectory (with $J=L+S$, $L$ intrinsic orbital angular momentum, $S$ total quark spin) have no mass-degenerate parity partner. We mention $f_2(1270)$, $f_4(2050)$, $f_6(2510)$, and $\Delta(1232)$, $\Delta(1950)7/2^+$, $\Delta(2420)11/2^+$. In \cite{Glozman:2010rp} it is argued that formation of the spin-parity partners of mesons on the leading Regge trajectory could be suppressed by orbital angular momentum barrier factors. We give two examples: in $\bar pp$ formation of $f_4(2050)$, $L=3$ is required; formation of its (unobserved or non-existing) mass-degenerate parity partner $\eta_4$ requires $L=4$. Likewise, $\Delta(1950)7/2^+$ requires $L=3$ in $\pi N$ scattering, and its $7/2^-$ parity partner $L=4$. Their non-observation could be assigned to their suppression by the angular-momentum barrier. This is different in photoproduction: a $7/2^+$ resonance needs an amplitude with $L=3$ between photon and nucleon, a $7/2^-$ resonance an amplitude with $L=2$. Photoproduction hence provides the best and possibly the only chance to find a decisive support or an experimental argument against the hypothesized restoration of chiral symmetry.

In this letter we report on a partial-wave analysis of the data on $\gamma p\to n\pi^+$ and $\gamma p\to p\pi^0$ covering differential cross sections $d\sigma/d\Omega$ \cite{vanPee:2007tw,Dugger:2007bt,Dugger:2009pn,Crede:2011dc}, the beam asymmetry $\Sigma$
\cite{Sparks:2010vb,Dugger:2013crn}, and the double-polarization observables
$T$, $P$, $H$ \cite{Hartmann:2014mya,Hartmann:2015kpa}, $G$ \cite{Thiel:2012yj,Thiel:2016chx},
and $E$ from the CLAS \cite{Strauch:2015zob} and CBELSA/TAPS \cite{Gottschall:2013uha} experiments. $\Sigma$, $T$, $P$, $H$ can be measured simultaneously when a transversely polarized target (with polarization $p_{T}$)  
and a linearly polarized photon beam (with polarization $p_{\gamma}$) are used:
\begin{align}
\frac{N(\phi)}{N_{0}} = 1 &- p_{\gamma}\Sigma_{\rm eff}\cos(2\phi) + p_{T} T \sin(\phi-\alpha) \nonumber \\[-1.5ex]
& - p_{T}\,p_{\gamma} P\cos(2\phi)\sin(\phi-\alpha) \nonumber \\
& + p_{T}\,p_{\gamma} H\sin(2\phi)\cos(\phi-\alpha),
\end{align}                                                                                                           
where $\phi$ is the azimuthal angle between the photon polarization plane and the scattering plane,
and $\alpha$ is the azimuthal angle between the target polarization vector and the photon polarization plane. 
$G$ can be deduced from the correlation between the photon polarization plane and the scattering plane
for protons polarized along the direction of the incoming photon; $E$ is defined by the
(normalized) difference between the cross sections for parallel and anti-parallel photon and
proton spin orientations. Data from older experiments \cite{GWU:2014now} are also included
in the partial wave analysis. The data are fitted
jointly with data on $N\eta$, $\Lambda K$, $\Sigma K$, $N\pi^0\pi^0$, and $N\pi^0\eta$ from both photo- and pion-induced reactions. Thus inelasticities in the meson-baryon system
are constrained by real data.  A list of the data used for the fit
can be found in
\cite{Anisovich:2011fc,Anisovich:2013vpa,Sokhoyan:2015fra,Gutz:2014wit} 
and on our website (pwa.hiskp.uni-bonn.de).

We shortly outline the analysis technique and fit strategy.
The helicity-dependent amplitude for photoproduction of the final
state $b$ is cast into the form \cite{Chung:1995dx}
\begin{eqnarray}
a^h_{b}&=&P^h_a(I-i\rho K)_{ab}^{-1}\qquad {\rm where}\\
P^h_a&=&\sum\limits_\alpha \frac{A^h_{\alpha}
g^{\alpha}_a}{M_\alpha^2-s}+F_{a}\,.\vspace{-2mm}
\end{eqnarray}
$A^h_{\alpha}$ is the photo-coupling of a pole $\alpha$
and $F_{a}$ a non-resonant transition. The helicity amplitudes $A^{1/2}_\alpha$, $A^{3/2}_\alpha$ are defined as residues of the helicity-dependent amplitude at the pole
position and are complex numbers. They become real and coincide with the conventional helicity amplitudes $A^{1/2}_{\rm BW}$, $A^{3/2}_{\rm BW}$, if a Breit-Wigner amplitude with constant width is used. For pion-induced reactions and $\pi N$ elastic scattering similar formulae apply.
The exact formulas used in the fits are documented in a series of papers
\cite{Anisovich:2004zz,Anisovich:2006bc,Anisovich:2007zz,Denisenko:2016ugz}.

\begin{figure}[pb]
\includegraphics[width=0.5\columnwidth]{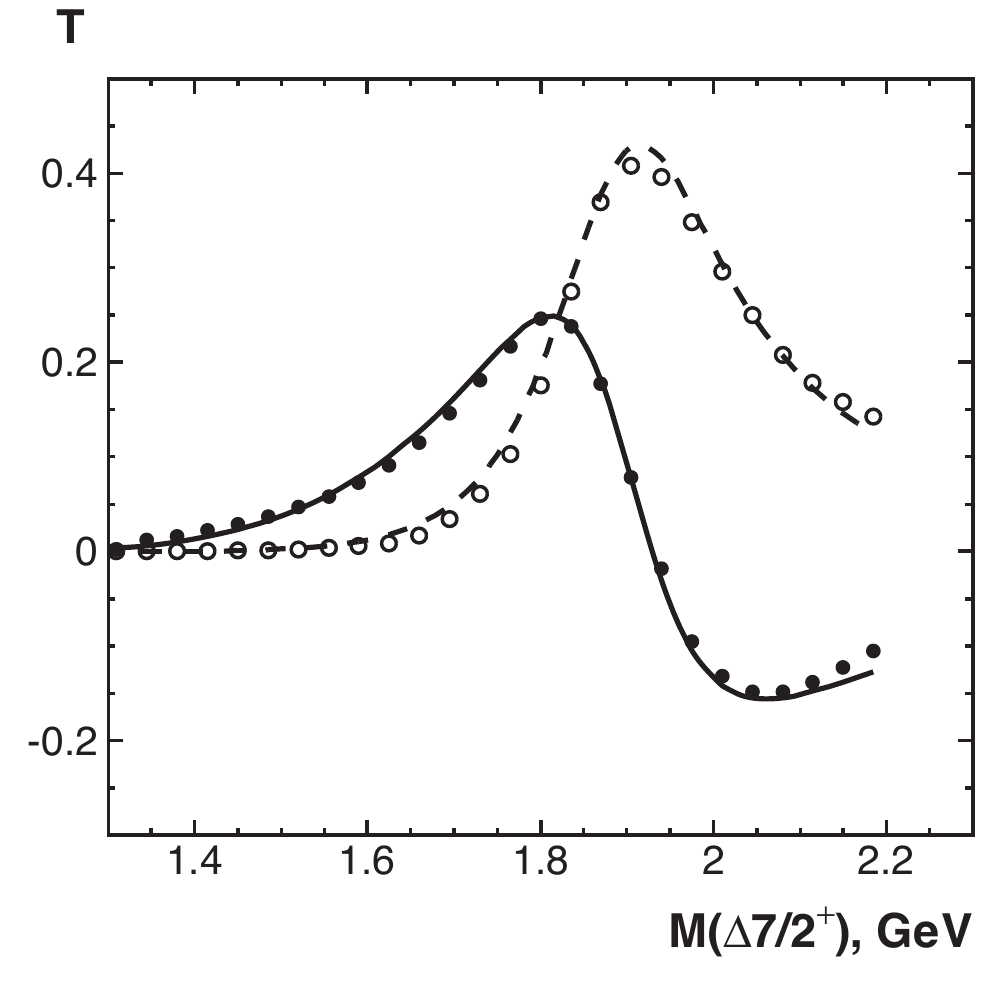}%
\includegraphics[width=0.5\columnwidth]{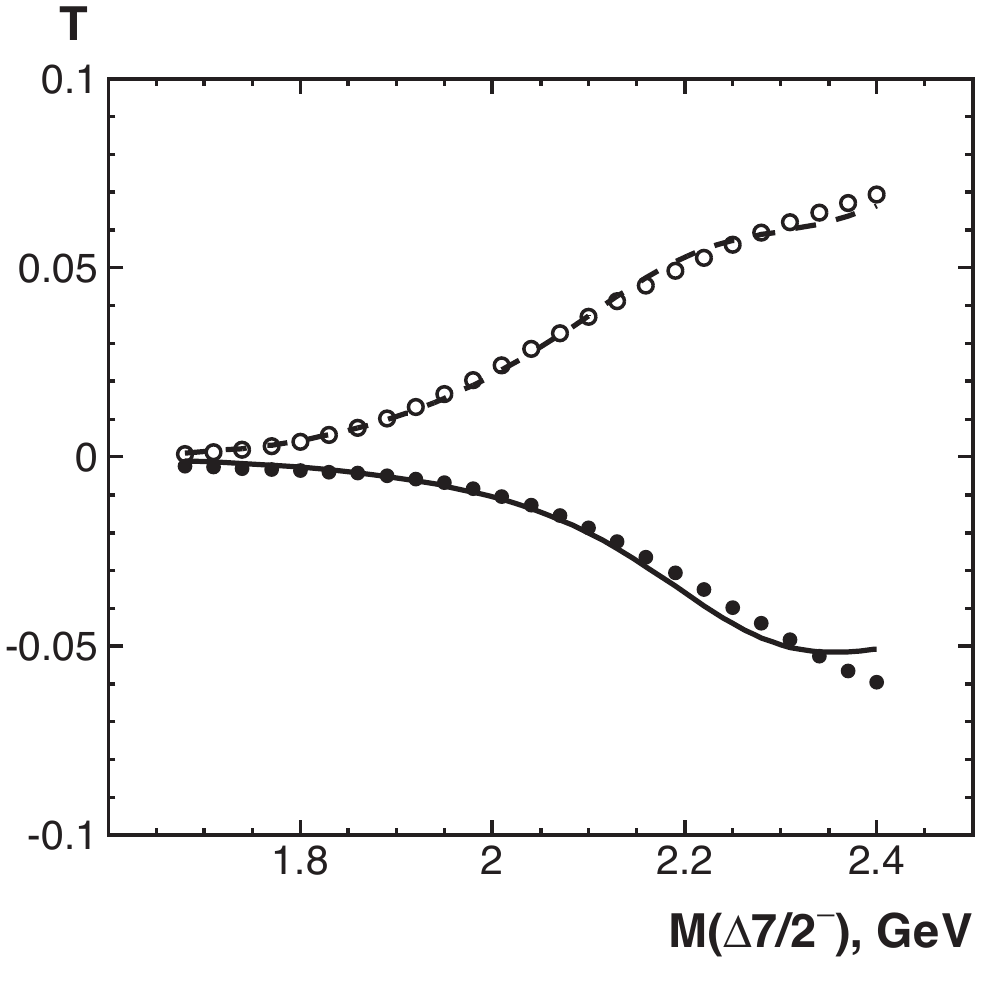}
\caption{\label{pwh}Real (solid points) and imaginary (open circles) part of the elastic $\pi N$ scattering amplitude $T$ in the $I(J^P)=3/2(7/2^+)$ and $3/2(7/2^-)$ partial waves \cite{Hohler:1984ux} and the BnGa fit. No uncertainties are known for \cite{Hohler:1984ux}. \vspace{-2mm}
}
\end{figure}

The $K$ matrix parametrizes resonances and background
contributions:\vspace{-1mm}
\begin{equation}
K_{ab}=\sum\limits_\alpha \frac{g^{\alpha}_a
g^{\alpha}_b}{M_\alpha^2-s}+f_{ab}\,.\vspace{-1mm}
\end{equation}
Here $g^{\alpha}_{a,b}$ are coupling constants of the pole $\alpha$
to the initial and the final state. The background terms $f_{ab}$
describe non-resonant transitions from the initial to the final
state. In most partial waves, a constant background term is sufficient to achieve a good fit.
Only the background in the meson-baryon $S$-wave required a more complicated form:\vspace{-2mm}
\begin{equation}
f_{ab}=\frac{(a+b\sqrt{s})}{(s-s_0)}.\vspace{-1mm}
\end{equation}
Further background contributions are obtained from the reggeized exchange of vector mesons \cite{Anisovich:2004zz}.

\begin{table*}[pt]
\caption{\label{BR}Properties of $\Delta(1950)7/2^+$ and $\Delta(2200)7/2^-$: masses M, widths 
$\Gamma$, helicity amplitudes $A_{1/2}$,  $A_{3/2}$ (in units of $10^{-3}$\,GeV$^{-1/2}$), 
and branching ratios 
BR$=\Gamma_i/\Gamma_{\rm tot}$. The RPP estimates \cite{Beringer:1900zz} are given in parenthesis. The fit yields the quantities $A_{1/2}\sqrt{\rm BR}$ and $A_{3/2}\sqrt{\rm BR}$. The branching ratio
for $N\pi$ decays from a fit to the $\pi N$ elastic scattering amplitude $T$  \cite{Hohler:1984ux} is
used to determine the helicity amplitudes $A_{1/2}$, $A_{3/2}$. Also listed are the decay orbital angular momenta $L$.}
\renewcommand{\arraystretch}{1.1}
\begin{center}
\begin{tabular}{ccccccccccc}
\hline\hline
\multicolumn{2}{c}{\boldmath$\Delta(1950)7/2^+$\quad ****} && \multicolumn{4}{c}{\boldmath$\Delta(2200)7/2^-$\quad *}\\
\hline
\multicolumn{1}{l}{$M =1917 \pm 4$ MeV} & \multicolumn{1}{l}{($1915 - 1950$ MeV)} && \multicolumn{4}{c}{$M = 2176 \pm 40$ MeV}\\
\multicolumn{1}{l}{$\Gamma = 251 \pm 8$ MeV} & \multicolumn{1}{l}{($235 - 335$ MeV)} && \multicolumn{4}{c}{$\Gamma = 210 \pm 70$ MeV}\\
\multicolumn{1}{l}{$A_{1/2} =-67 \pm 5$} & \multicolumn{1}{l}{($-76 \pm 12$)} && \multicolumn{2}{l}{$A_{1/2} = 60 \pm 20$}
&\multicolumn{2}{l}{$A_{3/2} = -(20 \pm 8)$} \\
\multicolumn{1}{l}{$A_{3/2} =-94 \pm 4$} & \multicolumn{1}{l}{($-97 \pm 10$)} &&Decay&$|A_{1/2}|\sqrt{\rm BR}$&$|A_{3/2}|\sqrt{\rm BR}$&BR \\
\hline
\multicolumn{1}{l}{$N\pi$, $L=3:$ \quad $46 \pm 2$\%} & \multicolumn{1}{l}{($35 - 45$\%)} && \multicolumn{1}{l}{$N\pi$, $L=4$} &$11\pm 5$&$4\pm 1.5$& $3.5 \pm 1.5$\%  \\
\multicolumn{1}{l}{$\Sigma K$, $L=3:$ \quad $0.6 \pm 0.2$\%} & \multicolumn{1}{l}{(seen)} && \multicolumn{1}{l}{$\Sigma K$, $L=4$} &$12 \pm 6$&$4 \pm 2$& $4 \pm 3$\%   \\
\multicolumn{1}{l}{$\Delta\pi$, $L=3:$ \quad $5 \pm 3$\%} & \multicolumn{1}{l}{($20 - 30$\%)}   && \multicolumn{1}{l}{$\Delta\pi$, $L=2$} &$50 \pm 8$&$17\pm 5$& $70 \pm 30$\%\\
\multicolumn{1}{l}{$\Delta\pi$, $L=5:$\quad --} & \multicolumn{1}{l}{(seen)}   && \multicolumn{1}{l}{$\Delta\pi$, $L=4$} &$23 \pm 15$&$8 \pm 5$& $15 \pm 10$\%\\
\multicolumn{1}{l}{$\Delta\eta$, $L=3:$ \quad $0.3 \pm 0.3$\%} & \multicolumn{1}{l}{}   && \multicolumn{1}{l}{$\Delta\eta$, $L=2$} &${\sim}5$&${\sim}2$& ${\sim}1$\%\\
\multicolumn{1}{l}{$\Delta\eta$, $L=5:$\quad --} & \multicolumn{1}{l}{}   && \multicolumn{1}{l}{$\Delta\eta$, $L=4$} &--&--&--\\
\hline\hline
\end{tabular}
\renewcommand{\arraystretch}{1.0}
\end{center}
\end{table*}
\begin{figure}[pt]
\vspace{-3mm}
\includegraphics[width=1.\columnwidth,height=0.52\textheight]{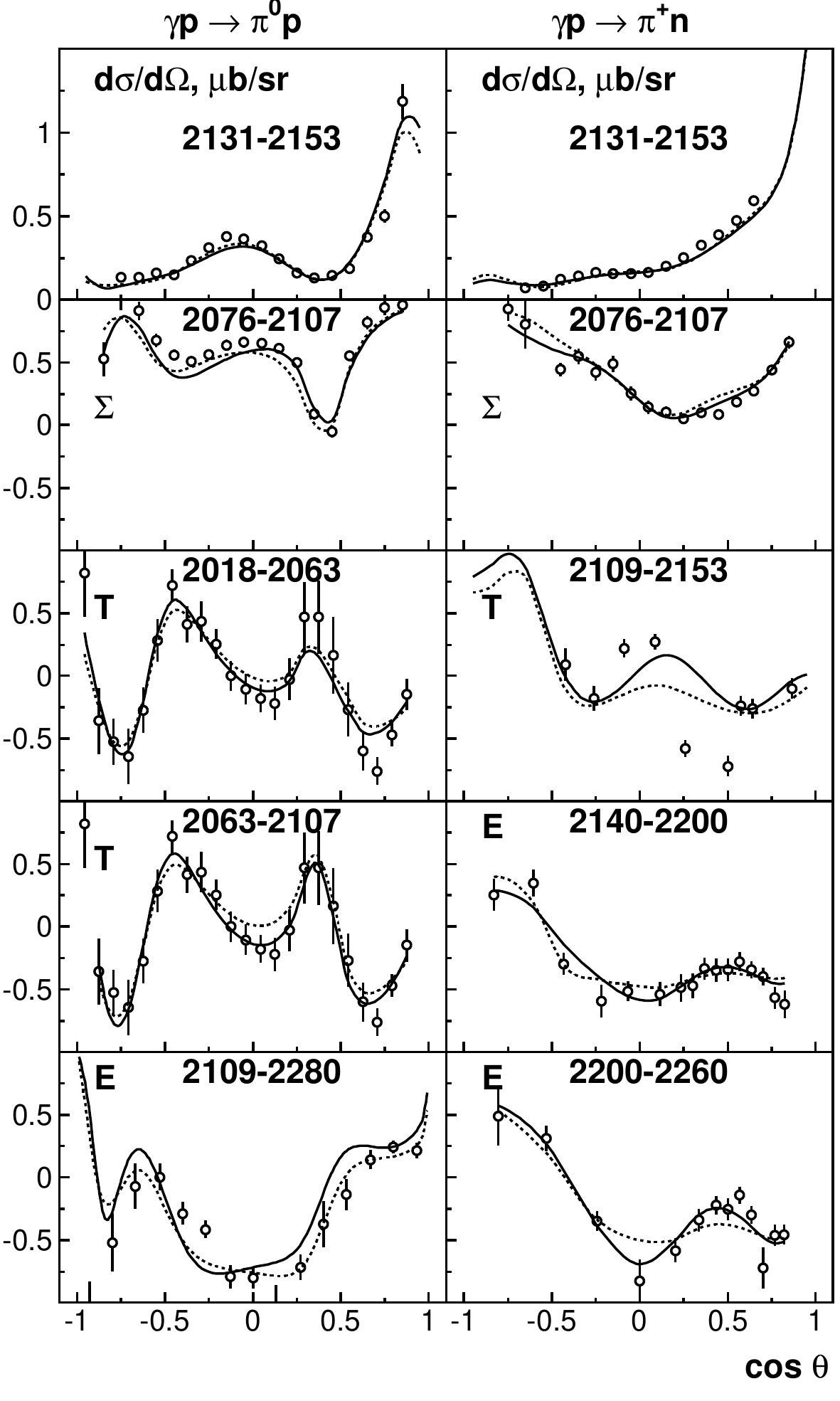}\vspace{-2mm}
\caption{\label{bestfit}Selected data and fits. Data: differential cross section $d\sigma/d\Omega$ \cite{Crede:2011dc,Dugger:2009pn}; target asymmetry $T$ \cite{Hartmann:2014mya,Hartmann:2015kpa,GWU:2014now}; beam asymmetry $\Sigma$ \cite{Dugger:2013crn}; helicity dependence $E$ \cite{Gottschall:2013uha,Strauch:2015zob}. Best fit: solid curve, fit without $\Delta(2200)7/2^-$: dotted curve. The fit deteriorated by $\delta\chi^2=597$ when $\Delta(2200)7/2^-$ was removed. The mass ranges are given in MeV. 
}
\end{figure}

To deduce branching ratios BR$=\Gamma_i/\Gamma_{\rm tot}$, a normalization is required. We use the $\pi N$ elastic scattering amplitudes $T$ determined by H\"ohler and collaborators \cite{Hohler:1984ux}, see Fig.~\ref{pwh}, which define the $\Delta(1950)7/2^+$ and $\Delta(2200)7/2^-\to \pi N$ branching ratios. From the fit to photoproduction data described below, the other branching ratios can be deduced.

We first made a fit to the data with our standard set of resonances and our standard fitting procedure using weights for low statistics data \cite{Anisovich:2011fc}. These contain all $N^*$ and $\Delta^*$ resonances with nominal masses up to 2.2~GeV listed in the Review of Particle Properties (RPP)~\cite{Beringer:1900zz} except $N(1685)$, $\Delta(1750)1/2^+$, $\Delta(2000)5/2^+$, $\Delta(2150)1/2^-$, for which we find no evidence. All masses, widths, and coupling constants
stay well within the errors quoted in \cite{Anisovich:2011fc}.
A few higher-mass reso\-nan\-ces are also taken into account (see \cite{Anisovich:2011fc}, for a list) which
improve the convergence behavior of the fit but their properties, in particular their masses and widths, remain undetermined. The large number of resonances offered to the fit avoids the problem that a group of unconsidered resonances could mimic a single resonance which then might lead to a significant improvement of the fit quality. With these ingredients, a reasonably good fit to all two-body reactions was achieved with a $\chi^2=49044$ for 32666 data points. These resonances are all included in the fit or were partly not used. The 
detailed results depend on model space which is used. 

The fits return properties of $\Delta(1950)7/2^+$ and $\Delta(2200)7/2^-$ (and of many other resonances). 
The results on $\Delta(1950)7/2^+$ and $\Delta(2200)7/2^-$ are listed in Table~\ref{BR}. Selected data 
and the fit curve are shown in Fig.~\ref{bestfit}. 
The values and the uncertainties in Table~\ref{BR} are derived from the mean values and 
from the spread of results of the various fits. Within 
the uncertainties, most properties of $\Delta(1950)7/2^+$ agree well with those reported in the 
RPP~\cite{Beringer:1900zz}. For $\Delta(2200)7/2^-$, RPP gives no estimates, a comparison with other determinations will be discussed below. The $\Delta(2200)7/2^-$ mass is compatible with its nominal mass, we assign an uncertainty of $\pm 40$~MeV.
The $\Delta(2200)7/2^-$ width is less well determined: several fits give $\Gamma\approx140$~MeV, other fits return $\Gamma\approx280$~MeV. Hence we quote $\Gamma= 210\pm 70$~MeV.

The comparison of the $N\pi$ branching ratios of $\Delta(1950)7/2^+$ and $\Delta(2200)7/2^-$ shows 
why $\Delta(1950)7/2^+$ is a well-established resonance and $\Delta(2200)7/2^-$ not: in elastic $\pi N$ 
scattering, the $N\pi$ branching ratio enters in the entrance and the exit channel, and the signature of 
$\Delta(2200)7/2^-$ in elastic scattering is more than 100 times weaker than that of $\Delta(1950)7/2^+$. 
Due to the weakness of the $\Delta(2200)7/2^-$ signal in elastic scattering, the $\Delta(2200)7/2^-\to N\pi$ 
branching ratio has a sizable uncertainty.  When this branching ratio is decreased by 20\%, all other 
branching ratios are reduced by 20\% while the squared helicity amplitudes increase by 20\%.

In the next step, we removed $\Delta(2200)7/2^-$ from the list of resonances used in the PWA, and the fit deteriorated visibly, see Fig.~\ref{bestfit}. We tried to replace $\Delta(2200)7/2^-$ by other resonances or by a group of up to three additional resonances. Numerous combinations of different quantum numbers were tested, but even the best-suited three resonances do not lead to the same improvement of the fit quality as $\Delta(2200)7/2^-$. When $\Delta(2200)7/2^-$ was then admitted as fourth additional resonance, the fit still improved significantly. Further, we excluded individual data sets on the polarization variables $\Sigma$, $T$, and $E$ 
for $\gamma p\to \pi^0p$ or $\gamma p\to \pi^+n$. Table~\ref{improve} gives  
the change in $\chi^2$ which we observe and which we should expect 
($N_{\rm events}\cdot 49044/32666$ with $N_{\rm events}$ being the number of
data points for a specified observable). The difference is assigned to the 
fractional evidence for the existence of $\Delta(2200)7/2^-$. Obviously,
all data sets contributed to the evidence for  
$\Delta(2200)7/2^-$. Most sensitive were the $\gamma p\to \pi^+n$ data on $E$ and $\Sigma$ and the   
$T$ and $E$ data on $\gamma p\to \pi^0p$. 
The six masses obtained in these fits deviate by $-(3\pm 10)$\,MeV, the widths by 
($1\pm13$)\,MeV from the mean value. These differences
are well covered by the spread of results obtained when different models were used to fit the data.

\begin{figure}[pt]
\includegraphics[width=1.\columnwidth,height=0.66\textheight]{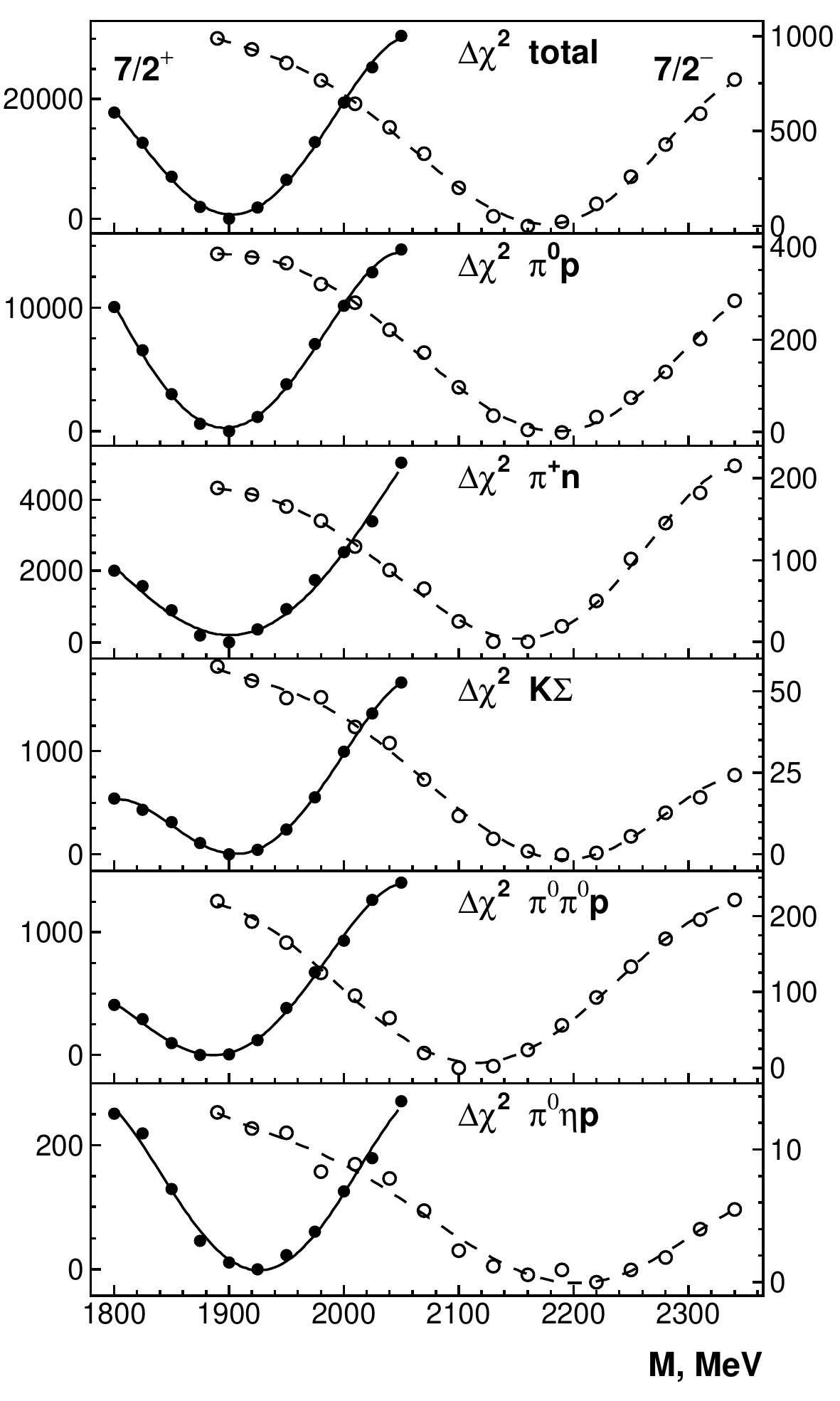}
\vspace{-5mm}
\caption{\label{scan-g37}The increase in pseudo-$\chi^2$ of the fit to a large body of pion- and photo-produced reactions when the mass of $\Delta(1950)7/2^+$ (solid points) or $\Delta(2200)7/2^-$ (open circles) is scanned. The scale on the left (right) abscissa refers to the $7/2^+$ ($7/2^-$) partial wave. The curves are to guide the eye.\vspace{-2mm}
}
\end{figure}
\begin{table}[pb]
\caption{\label{improve}The observed and the expected change in $\chi^2$ when single
data sets are removed from the fit. The differences are assigned to the fractional 
evidence for the existence of $\Delta(2200)7/2^-$.
}
{\footnotesize
\begin{center}
\renewcommand{\arraystretch}{1.4}
\begin{tabular}{ccccccc}
\hline\hline
\hspace{-2mm}Reaction\hspace{-2mm}                    & \multicolumn{3}{c}{$\gamma p\to \pi^0p$}& \multicolumn{3}{c}{$\gamma p\to \pi^+n$}\\
\hspace{-2mm}Observable\hspace{-2mm}&    $\Sigma$    &     T     &     E      &    $\Sigma$    &     T     &     E      \\
\hline
N$_{\rm event}$             &       702      &     609   &     509    &       389      &     661   &     140    \\
$\Delta\chi^2$ (obs.)   &      1346      &    1476   &    1372    &       1317     &    1497   &    1393    \\   
$\Delta\chi^2$ (exp.)   &      1054      &     915   &     764    &       584      &     992   &     210    \\   
\hline
\hspace{-2mm}$\Delta\chi^2_{\Delta(2200)7/2^-}$\hspace{-2mm}&292     &     541   &     608    &       733      &     505   &    1183    \\[1ex]   
\hline\hline
\end{tabular}
\end{center}
}
\end{table}

Next, we added data on three-body reactions, in particular $\gamma p\to \pi^0\pi^0 p$
\cite{Sokhoyan:2015fra} and $\gamma p\to \pi^0\eta p$ \cite{Gutz:2014wit}. The inclusion
of three-body reactions did not lead to any significant changes in mass, width, or
two-body couplings. Hence these values were frozen to their central values when the
three-body reactions were included. Three-body decays were assumed
to decay via intermediate meson or baryon resonances in an isobar ansatz.
The three-body data were fitted in an event-based likelihood method; hence no $\chi^2$ value
was returned by the fit. To quantify the improvement of a fit, we added to the change in
$\chi^2$ -- derived from the fit to two-body reactions -- the change in log likelihood
multiplied by 2. This number is referred to as pseudo-$\chi^2$. Since the absolute
value of the log-likelihood function is meaningless, we give only changes of the
pseudo-$\chi^2$.

With these data included, we performed mass scans in the $J^P=7/2^+$ and $J^P=7/2^-$ partial waves. In the mass scans, the $J^P=7/2^+$ and $J^P=7/2^-$ partial waves were not described by
$K$-matrix poles but represented by multichannel Breit-Wigner amplitudes, hence the optimal parameters for mass and width can differ. Figure~\ref{scan-g37} (top) shows the change of the resulting pseudo-$\chi^2$ as a function of the imposed mass of the $J^P=7/2^+$ or the $J^P=7/2^-$ resonance. The total pseudo-$\chi^2$ has clear minima at a mass of 1917~MeV for $J^P=7/2^+$ and 2176~MeV for $J^P=7/2^-$. When the masses are detuned
from the best values, the widths of the resonances become wide.

Figure~\ref{scan-g37} also shows a breakdown of the total pseudo-$\chi^2$ into contributions from specific reactions. Clear minima are observed in $\gamma p\to \pi^0\,p$, $\pi^+\,n$, $K\,\Sigma$, $\pi^0\pi^0 p$, and even in $\pi^0\eta p$ (due to $\Delta(1232)\eta$. The minima are found at 1913, 1917, 1922, 1904, 1942~MeV, respectively at
2186, 2155, 2193, 2115, 2200~MeV, consistent with the overall minima at 1917 and 2176~MeV. It is remarkable that the same minima which are found for $\Delta^*$ decays into $\pi N$ are as well seen in the other allowed decay channels. 

In spite of the small $\Delta(2200)7/2^-\to \pi N$ coupling, the largest evidence stems from photoproduction of single pions. There are two reasons: first, the highly constraining
polarization data and their statistical power define the angular-momentum decomposition
very well. Second, the sequential decays in $2\pi^0$ photoproduction allow for a large
flexibility in describing the data; hence the statistical significance of those data is reduced.

The scans (Fig.~\ref{scan-g37}) demonstrate clearly that the masses 
of $\Delta(1950)7/2^+$ and $\Delta(2200)7/2^-$
are different. We tried to include a second  $\Delta^*$  resonance with $7/2^-$ and a mass constrained to fall into the 1900 -- 1950~MeV mass range but the fit returned a vanishing amplitude for it. The difference in squared masses of the two resonances is $(1.06\pm 0.17)$~GeV$^2$, in excellent agreement with the slope of the leading Regge trajectory for $\Delta^*$'s of $(1.08\pm0.01)$~GeV$^2$.

To search for a mass-degenerate parity partner of $\Delta(1950)7/2^+$, 
 we did a series of fits trying to impose a $\Delta(1950)7/2^-$ with a mass restricted
in the range 1920 to 1980~MeV in addition to $\Delta(2200)7/2^-$. In all fits, both helicity amplitudes converged to
zero: there is no mass-degenerate parity partner of  $\Delta(1950)7/2^+$  in the data.

Evidence for the $\Delta(2200)7/2^-$ resonance has been reported before, see Table~\ref{old}.
The Review of Particle Properties lists it as a one-star resonance, the evidence was
considered as {\it poor}. $\Delta(2200)7/2^-$ is not seen in the elastic $\pi N$ scattering
analysis of the GWU group \cite{Workman:2012hx}; in the recent Bonn-GWU-J\"ulich analysis it
is included in the fits but the authors state that they {\it cannot claim much evidence}
either \cite{Ronchen:2014cna}.

\begin{table}[pt]
\caption{\label{old}(Breit-Wigner) mass, width (in MeV) and $\pi N$ decay branching ratio 
BR$=\Gamma_\pi/\Gamma_{\rm tot}$
of  $\Delta(2200)7/2^-$. Ref. \cite{Ronchen:2014cna} reports the pole position, no uncertainty is given.}
\begin{center}
\renewcommand{\arraystretch}{1.1}
\begin{tabular}{cccc}
\hline\hline
Mass& Width & BR &Ref\\
\hline
$2280\pm 80$ & $400\pm 150$&$9\pm 2$\% &\cite{Hendry:1978cd}\\
$2115\pm 60$ & $400\pm 100$&$5\pm 2$\%  &\cite{Hohler:1979yr}\\
$2200\pm 80$ & $450\pm 100$&$6\pm 2$\%  &\cite{Cutkosky:1980rh}\\
$2280\pm 40$ & $400\pm 50$ & &\cite{Candlin:1983cw}\\
2157        & 477        & &\cite{Ronchen:2014cna} \\
$\mathbf{2176\pm 40}$ & $\mathbf{210\pm 70}$ &$\mathbf{3.5\pm 1.5 \%}$ & this work\\
\hline\hline
\end{tabular}
\end{center}
\renewcommand{\arraystretch}{1.0}
\end{table}

Finally, we discuss the helicity couplings. The $\Delta(1950)7/2^+$ resonance can be excited by the
$E_{3+}$ and $M_{3+}$ multipoles, $\Delta(2200)7/2^-$ by the $E_{4-}$ and $M_{4-}$ multipoles. 
The multipoles are related to the helicity amplitudes by \cite{Workman:2013rca}
\begin{eqnarray}
\hspace{-5mm}A^{1/2}_{3+} =  -\frac 12\{5 E_{3+} +3 M_{3+} \};&\hspace{-5mm}&
A^{3/2}_{3+} =   \frac 12 \sqrt{15} \{ E_{3+} - M_{3+} \}\nonumber\\
\hspace{-5mm}A^{1/2}_{4-} =  -\frac 12\{4 E_{4-} -6 M_{4-} \};&\hspace{-5mm}&
A^{3/2}_{4-} =  -\frac 12 \sqrt{24} \{ E_{4-} + M_{4-} \}\nonumber
\end{eqnarray}
from which we deduce 
\begin{eqnarray}
\hspace{-5mm}\Delta(1950)7/2^+ & E_{3+} =&\hspace{-2mm} - (1.5\pm 1.5) 10^{-3}{\rm \,GeV}^{-1/2} \\ 
&M_{3+} =&\hspace{-1.8mm}  (47.1\pm 1.8) 10^{-3}{\rm \,GeV}^{-1/2}\\
\hspace{-5mm}\Delta(2200)7/2^- & E_{4-} =&\hspace{-2.5mm} - ( 7.1\pm \ 4.5) 10^{-3}{\rm \,GeV}^{-1/2} \\
&M_{4-} =&\hspace{-2mm} (15.3\pm 4.2) 10^{-3}{\rm \,GeV}^{-1/2}
\end{eqnarray}
In quark models, $\Delta(1950)7/2^+$ has a leading $(L,S)$ configuration $(L=2,S=3/2)$. When 
it is excited, the spin of one of the three quarks of the proton has to flip. This requires a
magnetic multipole. Indeed, the electric multipole $E_{3+}$ is much smaller than the magnetic
multipole $M_{3+}$. This observation can be compared with the small $E/M$ ratio of $\Delta(1232)$:
the photo-excitation of $\Delta(1232)$ requires a spin flip as well. 
The  leading $(L,S)$ configuration of $\Delta(2200)7/2^-$ is likely
$(L=3,S=1/2)$ \cite{Klempt:2012fy}. No spin flip is required and both, 
electric and magnetic multipoles, may contribute (not necessarily at the same 
strength). Within the errors, data are consistent with this conjecture.

Summarizing, we have reported strong evidence for the $\Delta(2200)7/2^-$ resonance from a coupled-channel analysis of a large data base. Mass, width, and decay branching ratios are determined.
The observed mass strongly favors quark models and AdS/QCD and is in conflict with models in which chiral symmetry is restored in the high-mass spectrum of meson and baryon resonances. The
analysis is based on the BnGa approach exploiting the energy-dependence of the photoproduction
amplitudes. There is the hope that the photoproduction multipoles with low orbital angular momenta might be determined in an energy-independent (and thus model-independent) analysis
when further polarization data become available. This would be the basis to derive the $\Delta(2200)7/2^-$ properties exploiting different model assumptions.

We thank the CLAS and CB-ELSA collaborations for letting us use their data prior to publication. E.K.~acknowledges stimulating discussions with L.~Glozman. This work is supported by the \textit{Deutsche Forschungsgemeinschaft} (SFB/TR16), the \textit{U.S. Department of Energy, U.S.  National Science
  Foundation}, and the \textit{Russian Science Foundation} (RSF 16-12-10267).





\end{document}